# Mass differences of SU(3) strange hadrons

Bernard F. Riley

NNC Ltd., Birchwood Park, Warrington WA3 6BZ, United Kingdom

bernard.riley@nnc.co.uk

Mass differences between octet strange baryons, octet and decuplet strange baryons, and nonet vector mesons and octet baryons of equal strangeness coincide (within the small experimental uncertainties) with the levels of a recently proposed hadron mass spectrum. A number, M, characterises the mass difference in relation to the strange quark mass of the spectrum. The mass difference of $\Lambda$ and $\Sigma^0$ ($\Delta I=1$) is characterised by M= +1, while the mass difference of $\Sigma^0$ and $\Sigma(1385)^0$ ($\Delta J=1$) is characterised by M=-1. Vector meson-baryon ($\Delta J=\frac{1}{2}$) mass differences are characterised by M=-1/2. The mass differences of strange decuplet baryon resonances ($\Delta S=1$) are characterised by M=-2 and M=-3, where M is equal to the strangeness of the more massive baryon.

The quark mass evaluations of the Particle Data Group, [1] are consistent with the values of 'mass levels' which descend from the Planck Mass (1.2210 ± 0.0009 x10$^{19}$ GeV), $m_P$, in a geometric series of common ratio $2/\pi$. [2] The mass of the nth level in the spectrum is defined as

$$m_n = \left(\frac{\pi}{2}\right)^{-n} m_P \qquad (1)$$

Hadron singlet states occupy 'higher order' mass sub-levels of fractional (half-integer, quarter-integer, eighth integer, etc) n; doublet states lie either side of mass levels. [2] Charged leptons, neutral vector meson resonances and the neutral vector boson, Z, occupy mass sub-levels in a remarkable symmetric arrangement. [3]

'Mass partnerships' occur between particles of, normally, equal strangeness but different isospin, isospin projection or spin, the particle mass differences approximating to the masses of principal (integer n) and low order levels. Where there is a difference in spin, the mass partnership is between isospin multiplets. For certain hadron mass partnerships, the mass differences calculated from Particle Data Group values, [4] are equal to the masses of principal levels, as shown in Figure 1. The mass difference of $\Lambda$ and $\Sigma^0$ is 76.959 ± 0.023 MeV, for which n = 103.001 ± 0.003, that of $\Sigma^0$ and $\Sigma(1385)^0$ is 191.1 ± 1.0 MeV, for which n = 100.99 ± 0.01, and that of $\Sigma^+$ and $\Sigma^-$ is 8.08 ± 0.08 MeV, for which n = 107.99 ± 0.02. In the mass partnership between K* (S=-1) and $\Sigma$ (S=-1), the mass representing the isospin doublet is that of the associated, interstitial, mass level;

the mass representing the isospin triplet is that of the neutral state. By this definition, the mass difference of K* and Σ is 298.85 ± 0.68 MeV, for which n = 99.996 ± 0.005. Similarly, the mass difference of ϕ (S=0, $s\bar{s}$) and Ξ (S=-2, $uss - dss$) is 298.13 ± 0.99 MeV, for which n = 100.002 ± 0.007. Uncertainties in doublet masses stem from uncertainty in $m_P$.

Many hadrons of identical spin-parity are also related in mass. [1] In such relationships, the mass of a hadron results from the addition of valence quark masses, as derived from the mass spectrum, to the mass of a neutral 'precursor' hadron. Examples of these 'mass construction' relationships are:

$$m_{K^{*0}}(896\,MeV) \approx m_{\rho^0} + m_d + m_s\,(897\,MeV) \tag{2}$$

$$m_{\phi}(1019\,MeV) \approx m_{\omega} + 2m_s\,(1024\,MeV) \tag{3}$$

$$m_{J/\psi}(3097\,MeV) \approx m_{\omega} + 2m_c\,(3095\,MeV) \tag{4}$$

$$m_{\Lambda_c^+}(2285\,MeV) \approx m_{\Lambda} + m_u + m_d + m_c\,(2283\,MeV) \tag{5}$$

$$m_{\Lambda_b^0}(5624 \pm 9\,MeV) \approx m_{\Lambda} + m_u + m_d + m_b\,(5609\,MeV) \tag{6}$$

In such relationships, the most massive hadron could be a singlet state, the most massive state of a doublet or the neutral state of a triplet. No exceptions to these rules have been identified. The mass construction relationships are not exact, at least for the singlet states, because the hadrons are constrained to occupy mass levels. [1]

Pairs of decuplet baryon resonances for which ΔS=1 are related in mass by a mass construction process in which fractional values of n correspond to the resonance mass differences. The mass difference of Δ(1232) and Σ(1385)⁰ is 151.5 ± 1.9 MeV,[1] for which n = 101.501 ± 0.028, that of Σ(1385)⁰ and Ξ(1530)⁻ is 151.3 ± 1.6 MeV, for which n = 101.504 ± 0.024, and that of Ξ(1530)⁰ and Ω⁻ is 140.65 ± 0.61 MeV, for which n = 101.665 ± 0.009.

---

[1] The Δ(1232) states are associated with a low order level, of mass 1232.2 MeV.

The values of mass difference (Δm) discussed here, between hadrons or multiplets of different isospin, spin or strangeness, fall within two principal mass level intervals of the strange quark level (n=102), of mass 120.92 MeV. Mass differences between strange hadrons are characterised by a value of $n_{\Delta m}$ which is related to the quantum numbers, I, J and S, of the particles. Where ΔI=1, $n_{\Delta m} - 102$ has the value +1. Where ΔJ=1, the quantity $n_{\Delta m} - 102$ has the value –1, and where ΔJ=½, $n_{\Delta m} - 102$ has the value –2. In the mass construction of Ξ⁻ (S=-2), $n_{\Delta m} - 102$ has the value -½, and in the mass construction of Ω⁻ (S=-3), $n_{\Delta m} - 102$ has the value $-\frac{1}{3}$. Defining the number, M, by

$$M = 1/(n_{\Delta m} - 102) \qquad (7)$$

values of this number for strange hadron mass differences are presented in Table 1 and Figure 2.

**Table 1      Values of M for strange hadron partnerships†**

| Hadron partnership | $n_{\Delta m}$ | $n_{\Delta m}$-102 | M=1/($n_{\Delta m}$-102) | Comments |
|---|---|---|---|---|
| Λ - Σ⁰ | 103.001 ± 0.003 | 1.001 ± 0.003 | 0.999 ± 0.003 | ΔI=1 |
| Σ⁰ - Σ(1385)⁰ | 100.99 ± 0.01 | -1.01 ± 0.01 | -0.99 ± 0.01 | ΔJ=1 |
| K* - Σ | 99.996 ± 0.005 | -2.004 ± 0.005 | -0.499 ± 0.001 | ΔJ=½ |
| φ - Ξ | 100.002 ± 0.007 | -1.998 ± 0.007 | -0.501 ± 0.001 | ΔJ=½ |
| Σ(1385)⁰ - Ξ(1530)⁻ | 101.504 ± 0.024 | -0.496 ± 0.024 | -2.02 ± 0.10 | S(Ξ⁻)=-2 |
| Ξ(1530)⁰ - Ω⁻ | 101.665 ± 0.009 | -0.335 ± 0.009 | -2.99 ± 0.08 | S(Ω⁻)=-3 |

† φ behaves as a strange meson with S=-2.

The values of M in Table 1 correspond closely in magnitude with the values of the quantum numbers associated with the mass differences. The values of strange hadron mass difference in Table 1 are related to M through

$$(\Delta m)_M = m_s \left(\frac{\pi}{2}\right)^{-\frac{1}{M}} \qquad (8)$$

A diagram showing the mass relationships existing between the hadrons and isospin multiplets of the SU(3) vector meson, baryon and baryon resonance multiplets is presented in Figure 3.

References

[1]  K. Hagiwara et al. (Particle Data Group), Phys. Rev. D **66**, 010001 (2002)

[2]  B. F. Riley, 'A unified model of particle mass', physics/0306098.

[3]  B. F. Riley, 'A geometric sequence of neutral vector boson and charged lepton masses', physics/0310138.

[4]  K. Hagiwara et al., Phys. Rev. D **66**, 010001 (2002) and 2003 off-year partial update for the 2004 edition available on the PDG WWW pages (URL: http://pdg.lbl.gov/)

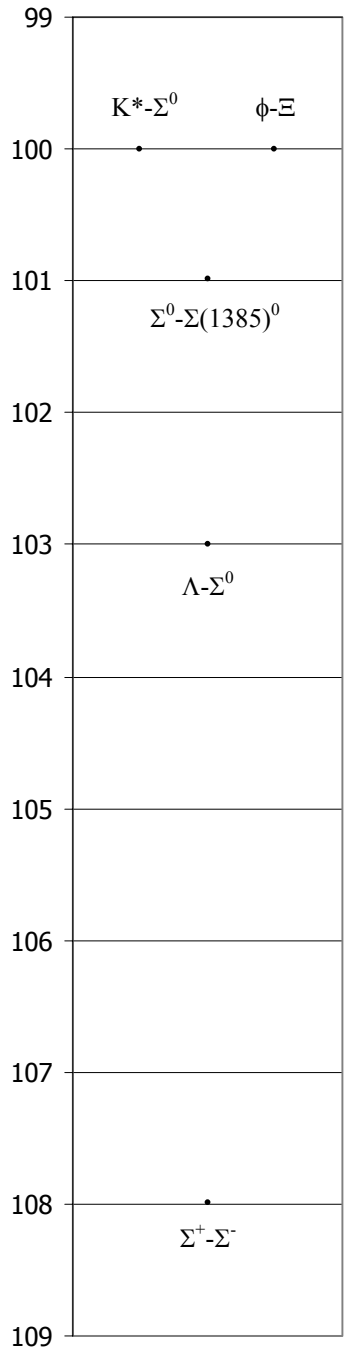

Figure 1: Values of n in (1), corresponding to strange hadron mass differences.

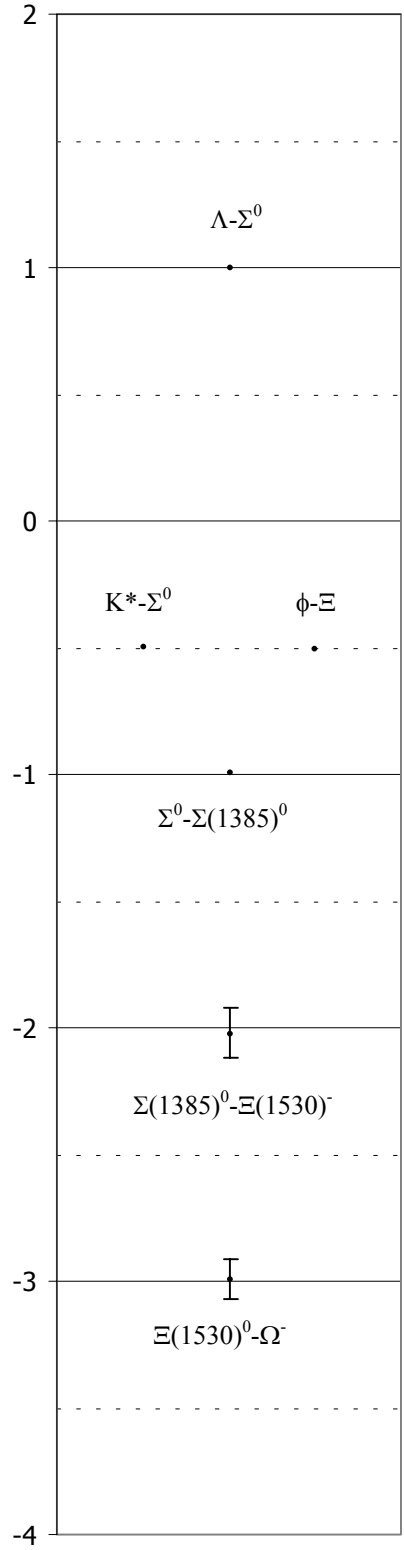

Figure 2: Values of M in (7), corresponding to strange hadron mass differences.

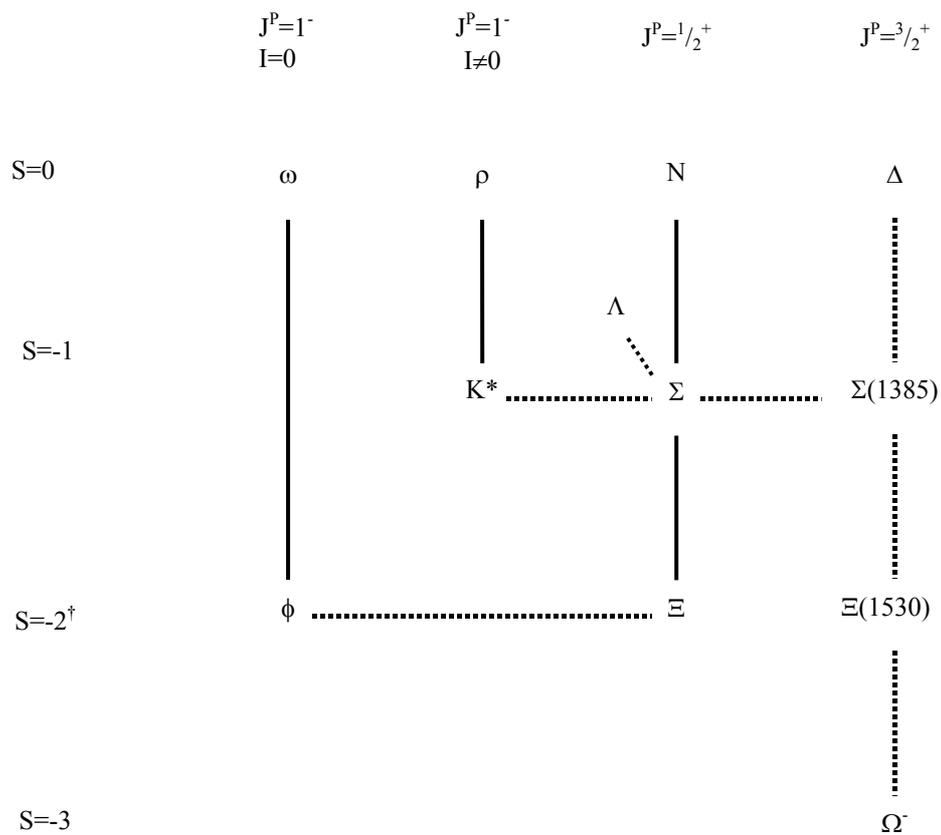

Figure 3: Mass relationships. Vertical lines join hadrons related through 'mass construction'. Dotted lines join hadrons (or multiplets) of mass difference characterised by integer or fractional values of n in (1).

† φ behaves as a strange meson with S=-2.